\documentstyle[12pt]{article}

\begin{document}
\setcounter{page}{1} \pagestyle{plain} \vspace{5cm}
\begin{center}
\large{\bf On the entanglement of formation of two-mode Gaussian states: a compact form}\\
\small
\vspace{1cm} {\bf Y. Akbari-Kourbolagh}\quad\ and \quad {\bf H. Alijanzadeh-Boura}\\
\vspace{0.5cm} {Department of Physics,
Azarbaijan Shahid Madani University,\\
Tabriz 53741-161, Iran\\
yakbari@azaruniv.edu,\, h.alijanzadeh@umz.ac.ir}
\end{center}
\vspace{1.5cm}
\begin{abstract}
We define an EPR-like uncertainty by using the Duan et al.'s
inequality which gives a necessary and sufficient condition for
the separability of any two-mode Gaussian state. We show that for
a given amount of entanglement, the uncertainty is minimized by
pure two-mode squeezed states. Using this fact, we write the
optimal pure-state decomposition and derive a compact form for the
entanglement of formation of two-mode Gaussian states. For
illustration  purposes, we consider symmetric and squeezed thermal
states as special cases and evaluate their entanglement of
formation explicitly. For the symmetric states, our result is in
agreement with the Giedke et al.'s one. To our knowledge, our work
is the first one which gives the exact entanglement of formation
of two-mode squeezed thermal states explicitly.

\end{abstract}
\newpage

\section{Introduction}

One of the main goals of quantum information science is to
quantify the entanglement or inseparability of quantum states. For
a pure bipartite state $ \vert \psi\rangle$, it is well known that
a convenient measure of entanglement is the von Neumann entropy
\cite{1,2,3}
$$E(\psi)=-{\mathrm{Tr}}(\rho_{A}\log_{2}\rho_{A})=-{\mathrm{Tr}}(\rho_{B}\log_{2}\rho_{B}),$$
where $\rho_{A}={\mathrm{Tr}}_{B}(\vert \psi\rangle\langle
\psi\vert)$ and $\rho_{B}={\mathrm{Tr}}_{A}(\vert
\psi\rangle\langle \psi\vert)$ are its reduced states. $E(\psi)$
is called the entropy of entanglement or the entanglement for
simplicity. However, there exists no unique measure of
entanglement in the case of mixed bipartite states and several
measures of entanglement have been introduced in this case
\cite{3a}.
\par
The entanglement of formation (EOF) is one of the measures with an
attractive physical motivation. For a bipartite mixed state,
Bennet et al. \cite{4} have defined this measure as the minimal
amount of average entanglement for any ensemble of bipartite pure
states realizing the state. Explicitly, the EOF of a mixed
bipartite state $ \rho $ is defined as
$$ E_{F}(\rho):=\inf\lbrace\sum_{k} p_{k} E(\vert\psi_{k}\rangle\langle\psi_{k}\vert):\quad \rho=
\sum_{k} p_{k} \vert\psi_{k}\rangle\langle\psi_{k}\vert\rbrace, $$
where the infimum is taken over all possible pure-state convex
decompositions of $ \rho $.
\par
In quantum information with continuous variables, Gaussian states
play an important role because they can be created relatively
easily and can be used in quantum cryptography and quantum
teleportation tasks \cite{5, 6, 7}. The first calculation of the
exact EOF in an infinite-dimensional Hilbert space has been
performed in the Giedke et al.'s remarkable work \cite{8} where
they have evaluated the exact EOF of a symmetric two-mode Gaussian
state by establishing a connection between its EPR-like
uncertainty and entanglement. In \cite{17}, Wolf et al. introduced
a Gaussian version of the EOF for bipartite Gaussian states by
considering merely their decompositions into pure Gaussian states
and showed that for symmetric two-mode Gaussian states, the
Gaussian EOF coincides with the exact EOF. In \cite{18}, Adesso et
al. computed Gaussian EOF for two important families of
nonsymmetric two-mode Gaussian states with extremal negativities
at fixed global and local purities. For an arbitrary two-mode
Gaussian state (TMGS), J. S. Ivan and R. Simon \cite{9} have
computed the EOF  based on a conjecture. Marians \cite{10} have
shown that the EOF of a TMGS coincides with its Gaussian EOF and
developed an insightful approach of evaluating the exact EOF.
Rigolin et al \cite{14} have derived two lower bounds on the EOF
of arbitrary mixed TMGSs. Oliveira et al. \cite{15} have
established tight upper and lower bounds for the EOF of an
arbitrary TMGS employing the necessary properties of Gaussian
channels. In the recent solution of the Gaussian optimizer
problem, Giovannetti et al. \cite{16} have computed the EOF for a
family of non-symmetric TMGSs and shown that it coincides with the
Gaussian EoF.
\par
Here, we derive a compact form for the EOF of arbitrary TMGS in
terms of four parameters which specify the standard form of its
covariance matrix (CM). We also obtain explicit results in the
special cases of symmetric TMGS and squeezed thermal states. For
the symmetric states, our result is in agreement with the Giedke
et al.'s one. To our knowledge, our work is the first one which
gives the exact entanglement of formation of two-mode squeezed
thermal states explicitly. To achieve this compact form, we define
an EPR-like uncertainty based on Duan et al.'s inequality
\cite{11}. Then we connect the entanglement of pure states with
their EPR-like uncertainties and show that pure two-mode squeezed
states are the least entangled states for a given uncertainty of
this type. Ultimately, we show that the pure-state decomposition
which leads to the EOF is established by a two-mode squeezed state
together with its displaced versions. The advantage of our work is
that it reduces the evaluation of EOF for the TMGSs to the
solution of two simple algebraic equations and provides a simple
method for determining the exact EOF of any TMGS.
\par
This paper is structured as follows: In Section 2 we provide a
brief description of the two-mode Gaussian states, including the
standard form of their covariance matrices and Duan et al.'s
inequality. In Section 3 we derive a compact form for the EOF of
an arbitrary TMGS. In Section 4 we present two examples to
illustrate the topic and compare our results with other ones. The
paper is ended with a brief conclusion in section 5.
\section{Gaussian states}
Let us consider a Gaussian state $\sigma$ of two single modes $A$
and $B$ described by the amplitude operators
$\hat{a}_{A}=\frac{\hat{x}_{A}+i \hat{p}_{A}}{\sqrt{2}}$ and
$\hat{a}_{B}=\frac{\hat{x}_{B}+i \hat{p}_{B}}{\sqrt{2}}$,
respectively, in which the canonical quadrature operators
$\hat{x}_{k}$, $\hat{p}_{l}$ have the commutators
$[\hat{x}_{k},\hat{p}_{l}]=i\,\delta_{kl}$ for all $k,l=A,B$.
Introducing the notation
$\hat{\xi}=(\hat{x}_{A},\,\hat{p}_{A},\,\hat{x}_{B},\,\hat{p}_{B})$,
the commutators can be expressed as
\begin{equation}\label{}
    [\hat{\xi}_{k},\hat{\xi}_{l}]=i\Omega_{k,l}\textsl{I}\quad,\quad
    k,l=1,2,3,4
\end{equation}
where $\textsl{I}$ is the identity operator and
$$\Omega=\bigoplus_{i=1}^{2}J\quad \textsl{with}\quad J=\left(
\begin{array}{cc}
0 & 1 \\
-1 & 0
\end{array}
\right).$$\\The state $\sigma$ can also be specified by its
characteristic function (CF)
\begin{equation}\label{}
    \chi_{\sigma}(\xi)={\mathrm{Tr}}(\sigma \hat{D}(\xi)),
\end{equation}
where $\xi=(x_{A},\,p_{A},\,x_{B},\,p_{B})^{T}$ is a real vector
and $\hat{D}(\xi)$ is a two-mode Weyl displacement operator
\begin{equation}\label{}
    \hat{D}(\xi)=\exp(i\xi^{T}\Omega\;\hat{\xi}).
\end{equation}
in which $T$ means transposition. The CF of TMGS $\sigma$ has the
following form
\begin{equation}\label{}
    \chi_{\sigma}(\xi)=\exp(i\xi^{T}\Omega
    d-\frac{1}{4}\xi^{T}\Omega^{T}\gamma_{\sigma}\Omega\xi),
\end{equation}
in which $d$ is its real displacement vector and $\gamma_{\sigma}$ is its
covariance matrix (CM) defined by
\begin{equation}\label{}
    d_{k}={\mathrm{Tr}}(\hat{\xi}_{k}\sigma)\quad,\quad
    \gamma_{{\sigma}_{k,l}}={\mathrm{Tr}}[(\hat{\xi}_{k}\hat{\xi}_{l}+\hat{\xi}_{l}\hat{\xi}_{k})\sigma]
    -2{\mathrm{Tr}}(\hat{\xi}_{k}\sigma)\,{\mathrm{Tr}}(\hat{\xi}_{l}\sigma).
\end{equation}
By definition, the CM $\gamma$ is a real, symmetric and positive
$4\times 4$ matrix. It turns out that a matrix $\gamma$ is a bona
fide CM iff it satisfies the uncertainty relation \cite{12}
\begin{equation}\label{e16}
\gamma+i\Omega\geq 0.
\end{equation}
Note that the displacement vector $d$ can always be shifted to
zero by a sequence of local unitary operations. Hence, it is
irrelevant for the study of entanglement and without lose of
generality we may take it to be zero and work only with TMGS of
zero displacement vector which is completely characterized by its
CM.
\par
In lemma 2 of \cite{11} it has been shown that the CM of a TMGS
can be transformed to the following standard form
\begin{equation} \label{v2}
 \gamma=\left(
\begin{array}{cccc}
nr_{1} & 0 & \sqrt{r_{1}r_{2}}k_{x}& 0 \\
 0 & n/r_{1} & 0  & k_{p}/\sqrt{r_{1}r_{2}} \\
\sqrt{r_{1}r_{2}}k_{x} & 0  & mr_{2} & 0  \\
 0 &k_{p}/\sqrt{r_{1}r_{2}} & 0  & m/r_{2}
\end{array}
\right).
\end{equation}
Here $ r_{1} $, $ r_{2} $ are arbitrary non-negative one-mode
squeezing factors, $ n,m\geq 1 $, $ k_{x}\geq -k_{p} > 0 $ and we have
\begin{equation}\label{ca}
\frac{nr_{1}-1}{mr_{2}-1}=\frac{n/r_{1}-1}{m/r_{2}-1},
\end{equation}
\begin{equation} \label{cb}
\vert \sqrt{r_{1}r_{2}}k_{x}\vert -\vert
k_{p}/\sqrt{r_{1}r_{2}}\vert
=\sqrt{(nr_{1}-1)(mr_{2}-1)}-\sqrt{(n/r_{1}-1)(m/r_{2}-1)}.
\end{equation}
It has been proved that Eqs. (\ref{ca}) and (\ref{cb}) have at
least one solution $ r_{1} $, $ r_{2}\geq 1 $ for a given set of
parameters $n,m,k_{x}$ and $k_{p}$ \cite{11}. Therefore, the CM of a
Gaussian state can be completely described by these four
parameters. For example, a standard two-mode squeezed vacuum state
$\vert\psi_{r}\rangle $ with squeezing parameter $r> 0$ is a TMGS
for which $r_{1}=r_{2}=1$ and the CM parameters are $n=m=\cosh2r$ and
$k_{x}=-k_{p}=\sinh2r$.
\par
For a given bipartite pure state $\vert \psi\rangle$, let us write
its Schmidt decomposition as
\begin{equation}\label{e9}
\vert \psi\rangle=\sum_{N=0}^{\infty}c_{N}\vert
u_{N}\rangle_{A}\otimes\vert v_{N}\rangle_{B},
\end{equation}
where $ c=(c_{0}, c_{1}, ...)$ is the set of non-negative Schmidt
coefficients in decreasing order with $\Vert c\Vert^{2}:=
\sum_{N=0}^{\infty}c_{N}^{2}=1 $ and $ \lbrace \vert
u_{N}\rangle_{A}\rbrace $ and $ \lbrace \vert
v_{N}\rangle_{B}\rbrace $ are orthonormal bases in the Hilbert spaces
of modes $A$ and $B$, respectively. Then the entropy of
entanglement is expressed as
\begin{equation}\label{e4a}
 E(\psi)=e(c):=-\sum_{N=0}^{\infty}c_{N}^{2}\log_{2}(c_{N}^{2}).
\end{equation}
For example, the state $\vert\psi_{r}\rangle$ has the Schmidt form
\cite{8}
\begin{equation}\label{e4}
\vert\psi_{r}\rangle=\sum_{N=0}^{\infty} c_{N}\vert N\rangle_{A}
\otimes \vert N\rangle_{B}\quad,\quad c_{N}=\frac{ \tanh^{N}r}{
\cosh r},
\end{equation}
where $\{\vert N\rangle_{A}\}$ ($\{\vert N\rangle_{B}\}$) is the
standard Fock basis for the mode $A$ ($B$) such that
$\hat{a}_{A}^{\dag}\hat{a}_{A}\vert N\rangle_{A}=N\vert
N\rangle_{A}$ and $\hat{a}_{B}^{\dag}\hat{a}_{B}\vert
N\rangle_{B}=N\vert N\rangle_{B}$. The entropy of entanglement for
$\vert\psi_{r}\rangle$ is calculated to be
\begin{equation}\label{e5}
 E(\psi_{r})=\cosh^{2}r \log_{2}(\cosh^{2}r)-\sinh^{2}r \log_{2}(\sinh^{2}r).
\end{equation}
\par
Duan et al. \cite{11} have shown that any bipartite separable
quantum state $ \varrho $ satisfies the inequality
\begin{equation}\label{e1}
\langle(\Delta\hat{u})^{2}\rangle_{\varrho}+\langle(\Delta\hat{v})^{2}\rangle_{\varrho}\,\geq
a^{2}+\frac{1}{a^{2}},
\end{equation}
where $\hat{u}=|a|\,\hat{x}_{A}+\frac{\hat{x}_{B}}{a}$ and
$\hat{v}=|a|\,\hat{p}_{A}-\frac{\hat{p}_{B}}{a}$ are EPR-like
operators, $ a $ is a real nonzero parameter and
$\langle(\Delta\hat{u})^{2}\rangle_{\varrho}={\mathrm{Tr}}(\varrho\hat{u}^{2})-[{\mathrm{Tr}}(\varrho\hat{u})]^{2}$
is the variance of $\hat{u}$. Furthermore, in proposition 2 they
have proved that inequality (\ref{e1}) is a necessary and
sufficient condition for the separability of TMGSs provided that $
a=-a_{0} $ with
$a_{0}^{2}=\sqrt{\frac{mr_{2}-1}{nr_{1}-1}}=\sqrt{\frac{m/r_{2}-1}{n/r_{1}-1}}$.
For inseparable states, however, the uncertainty relation
$\langle(\Delta\hat{u})^{2}\rangle_{\varrho}+\langle(\Delta\hat{v})^{2}\rangle_{\varrho}\,\geq
|\langle[\hat{u},\hat{v}]\rangle_{\varrho}|$ requires that the
total variance of operators $\hat{u}$ and $\hat{v}$ to be larger
than or equal to $|a^{2}-\frac{1}{a^{2}}|$, which reduces to zero
for the case $|a|=1$.
\section{Entanglement of formation of Gaussian states}
For any bipartite quantum state $\rho$, inequality (\ref{e1})
allows us to define an EPR-like uncertainty as follows
\begin{equation}\label{e2}
\Delta(\rho):=\min\lbrace 1,
\frac{\langle(\Delta\hat{u})^{2}\rangle_{\rho}+\langle(\Delta\hat{v})^{2}\rangle_{\rho}}{a^{2}+\frac{1}{a^{2}}}\rbrace.
\end{equation}
It is obvious from the definition and the foregoing facts that $\Delta(\rho)\in [b,1]$ with $b:=|a^{2}-\frac{1}{a^{2}}|/(a^{2}+\frac{1}{a^{2}})=\sqrt{1-\frac{4}{(a^{2}+\frac{1}{a^{2}})^{2}}}$.
By inequality (\ref{e1}), $b\leq\Delta(\rho)\,<1 $ is an evidence
for the inseparability of $\rho$. It is interesting to note that by
introducing
$$\sin\theta:=\frac{|a|}{\sqrt{a^{2}+\frac{1}{a^{2}}}}\quad,\quad \cos\theta:=\frac{1}{|a|\sqrt{a^{2}+\frac{1}{a^{2}}}}, $$
our definition of $\Delta(\rho)$ in Eq. (\ref{e2}) takes the same
form of the generalized EPR correlation $\Lambda_{\theta}$
introduced in \cite{9}.
\par
For a bipartite pure state $|\psi\rangle$ with
$b\leq\Delta(\psi)<1$ and with Schmidt decomposition (\ref{e9}),
Eq. (\ref{e2}) gives
\begin{equation}\label{e11}
  \begin{array}{c}
     \Delta(\psi)=1+\frac{2a^{4}}{1+a^{4}}\sum_{N=0}^{\infty}c_{N}^{2}\langle u_{N}\vert \hat{a}_{A}^{\dagger}\hat{a}_{A} \vert u_{N}\rangle
+\frac{2}{1+a^{4}}\sum_{N=0}^{\infty}c_{N}^{2}\langle v_{N}\vert
\hat{a}_{B}^{\dagger}\hat{a}_{B} \vert v_{N}\rangle \\
       -\frac{2}{a^{2}+\frac{1}{a^{2}}}\sum_{N,
M=0}^{\infty}c_{N}c_{M}\lbrace \langle u_{M}\vert \hat{a}_{A}
\vert u_{N}\rangle \langle v_{M}\vert \hat{a}_{B}\vert
v_{N}\rangle + c.c.\rbrace. \\
  \end{array}
\end{equation}
Also for a TMGS $\sigma$ for which the displacement vector is zero
and the CM is given by Eq. (\ref{v2}), we have
\begin{equation}\label{efa}
\Delta(\sigma)=\min\{ 1,
\frac{a^{2}\frac{nr_{1}+n/r_{1}}{2}+\frac{mr_{2}+m/r_{2}}{2a^{2}}+\frac{|a|}{a}(\sqrt{r_{1}r_{2}}k_{x}
-\frac{k_{p}}{\sqrt{r_{1}r_{2}}})}{a^{2}+\frac{1}{a^{2}}}\}.
\end{equation}
In the case of $\vert\psi_{r}\rangle$ this reduces to
\begin{equation}\label{e6}
\Delta (\psi_{r})=\min\{ 1, \cosh(2r)+ \frac{2 \vert
a\vert}{a}\frac{\sinh(2r)}{a^{2}+\frac{1}{a^{2}}} \}.
\end{equation}
This reveals the entanglement of $\vert\psi_{r}\rangle$ when $
\Delta (\psi_{r})\in [b, 1) $, i.e.,
 \begin{equation}\label{e7}
 0 < \tanh r  < -2\frac{\vert
 a\vert}{a}\frac{1}{a^{2}+\frac{1}{a^{2}}}\;.
  \end{equation}
Since $0 < \tanh r<1$, Eq. (\ref{e7}) implies that $ a < 0 $.
Therefor, we have
\begin{equation}\label{e7a}
    \Delta (\psi_{r})=\cosh(2r)-\frac{2}{a^{2}+\frac{1}{a^{2}}}\sinh(2r)\quad,\quad
    a<0 \quad,\quad \tanh r  < \frac{2}{a^{2}+\frac{1}{a^{2}}}.
\end{equation}
The EPR-like uncertainty $\Delta (\psi_{r})$ has the minimum value
$b$. Hereafter, we
consider only the case of $a<0$.
\par
For a fixed value of the parameter $ a $, any value of
$\Delta\in[b,1)$ can be achieved by a two-mode squeezed state. To
show this fact, we set $\Delta(\psi_{r})=\Delta$ in Eq.
(\ref{e7a}) and solve it for $r$. It is easy to see that this
equation has two solutions for $r$ provided that $\Delta\geq b$.
To determine which solution gives the desired two-mode squeezed
state, we plot $\Delta(\psi_{r})$ versus $r$ in the range $(0,
\tanh ^{-1}\frac{2}{a^{2}+1/a^{2}}) $ (Figure 1). As the figure
shows, for small $ r $, $\Delta(\psi_{r})$ decreases with $ r $,
whereas for large $ r $, it increases. We will show in the Lemma 2
below that the smaller solution has the right behavior and hence
it is the relevant one. Denoting this solution by $r_{\Delta'}$,
we have:
\begin{equation}\label{e8}
e^{-2r_{\Delta'}}=\frac{\Delta+\sqrt{\Delta^{2}-b^{2}}}{1+\sqrt{1-b^{2}}}:=\Delta'.
\end{equation}
\begin{figure}[htp]
\begin{center}\includegraphics{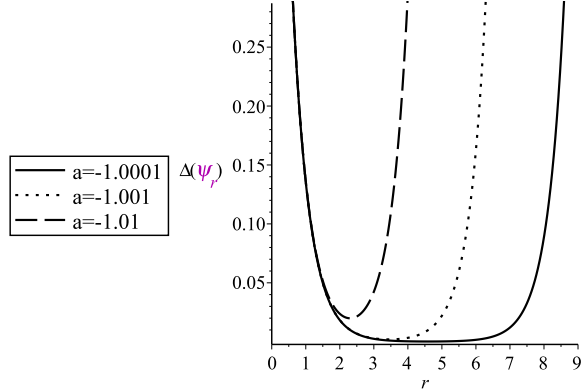} \vspace{6cm}
\end{center}
\caption{\small {$ \Delta(\psi_{r}) $ in terms of $ r $ for various values of $ a $.}}
\end{figure}
\par
Given the above definitions and facts, we are now in a position to
state the following proposition which is a generalized form of the
proposition 1 of Giedke et al.'s work \cite{8}.
\par {\bf{Proposition 1:}} For all pure states $|\psi\rangle$ of a two-mode system with $\Delta(\psi)\in[b,1]$
we have $ E(\psi)\geq E \left(\psi_{r_{\Delta'(\psi)}}\right)$,
where
\begin{equation}\label{}
  \Delta'(\psi):=\frac{\Delta(\psi)+\sqrt{\Delta^{2}(\psi)-b^{2}}}{1+\sqrt{1-b^{2}}}\quad,\quad
  e^{-2r_{\Delta'(\psi)}}:=\Delta'(\psi).
\end{equation}
\par Like Giedke et al., we prove this proposition by two lemmas and one definition. For a given $c=(c_{0}, c_{1}, ...)$, we define
\begin{equation}\label{e10}
\delta(c)= 1+2\sum_{N=0}^{\infty}(c_{N}^{2}-\frac{2}{a^{2}+\frac{1}{a^{2}}}\;c_{N}c_{N-1})N.
\end{equation}
It is obvious that $ \delta(c)\leq 1$ when $c_{N}\leq
\frac{2}{a^{2}+1/a^{2}}c_{N-1}$ for all $N$, and by Eq.
(\ref{e11}) we have $ \delta(c)=\Delta(\psi)$ whenever in Eq.
(\ref{e9}) the Schmidt basis coincides with Fock basis.
\par
{\bf{ Lemma 1}:} For all $|\psi\rangle$ with Schmidt decomposition
(\ref{e9}) which have the same set of Schmidt coefficients
satisfying the constraints $c_{N}\leq
\frac{2}{a^{2}+1/a^{2}}c_{N-1}$ for all $N$, we have $
\Delta(\psi) \geq \delta(c) $ and hence $\Delta'(\psi) \geq
\delta'(c)$, where
\begin{equation}\label{}
  \delta'(c):=\frac{\delta(c)+\sqrt{\delta^{2}(c)-b^{2}}}{1+\sqrt{1-b^{2}}}.
\end{equation}
\par
\textbf{Proof:} Since in this case $ \delta(c)\leq 1 $ we only
need to consider $|\psi\rangle$ with $b\leq\Delta(\psi) < 1 $. From
Eq. (\ref{e11}), we note that $ \Delta(\psi)\geq \min\lbrace Z(u),
Z(v)\rbrace:=Z $, where
\begin{equation}\label{e12}
Z=1+2\sum_{N=0}^{\infty}c_{N}^{2} \langle u_{N}\vert
\hat{a}_{A}^{\dagger}\hat{a}_{A}\vert u_{N}\rangle
-\frac{4}{a^{2}+\frac{1}{a^{2}}}\sum_{N,
M=0}^{\infty}c_{N}c_{M}\vert \langle u_{N}\vert
\hat{a}_{A}^{\dagger}\vert u_{M}\rangle \vert ^{2}.
\end{equation}
Let us now define $P:=\sum_{N=0}^{\infty}c_{N}^{2}\vert u_{N}\rangle\langle u_{N}\vert$ which is a density operator
with eigenvalues $c_{N}^{2}$ and eigenvectors $\vert u_{N}\rangle$. By this, Eq. (\ref{e12}) takes the form
\begin{equation}\label{e12a}
Z=1+2\;{\mathrm{Tr}}(P\hat{a}_{A}^{\dagger}\hat{a}_{A})
-\frac{4}{a^{2}+\frac{1}{a^{2}}}{\mathrm{Tr}}(\sqrt{P}\;\hat{a}_{A}^{\dagger}\sqrt{P}\;\hat{a}_{A}).
\end{equation}
Since the trace operation is basis independent, we take the first trace in the Fock basis and get
\begin{equation}\label{e12b}
Z=1+2\sum_{N=0}^{\infty}N \langle N\vert
P\vert N\rangle
-\frac{4}{a^{2}+\frac{1}{a^{2}}}{\mathrm{Tr}}(\sqrt{P}\;\hat{a}_{A}^{\dagger}\sqrt{P}\;\hat{a}_{A}).
\end{equation}
Our aim is to show that $Z \geq\delta(c)$, i. e.
\begin{equation}\label{e12c}
\begin{array}{c}
  1+2\sum_{N=0}^{\infty}N \langle N\vert
P\vert N\rangle -\frac{4}{a^{2}+\frac{1}{a^{2}}}{\mathrm{Tr}}(\sqrt{P}\;\hat{a}_{A}^{\dagger}\sqrt{P}\;\hat{a}_{A})\\
  \geq 1+2\sum_{N=0}^{\infty}(c_{N}^{2}-\frac{2}{a^{2}+\frac{1}{a^{2}}}\;c_{N}c_{N-1})N.
\end{array}
\end{equation}
This inequality can be rewritten as
\begin{equation}\label{e12d}
\frac{2}{a^{2}+\frac{1}{a^{2}}}\left({\mathrm{Tr}}(\sqrt{P}\;\hat{a}_{A}^{\dagger}\sqrt{P}
\;\hat{a}_{A})-\sum_{N=0}^{\infty}c_{N}c_{N-1}N\right)\leq \sum_{N=0}^{\infty}N \langle N\vert
P\vert N\rangle-\sum_{N=0}^{\infty}c_{N}^{2}N.
\end{equation}
As it was shown in \cite{8}, the inequality (\ref{e12d}) is valid for the case of $a=-1$. Since the factor
 $2/(a^{2}+\frac{1}{a^{2}})$ is positive and less than or equal to one, it is obvious that
the inequality will also be valid for other values of $a$ provided
that we prove the non-negativity of the right hand side. For this
purpose, we use the Schur-Horn theorem \cite{19,20}. Based on this
theorem, if $\mathcal{A}$ be a self-adjoint
 operator in a finite dimensional Hilbert space or a positive compact operator in an infinite-dimensional Hilbert space
with eigenvalues vector $\lambda({\mathcal{A}})=(\lambda_{1}\geq\lambda_{2}\geq\cdots)$ and
if $\mathrm{diag}({\mathcal{A}})=(\mu_{1}\geq\mu_{2}\geq\cdots)$ denotes a vector whose components are
the diagonal entries of $\mathcal{A}$ with respect to some orthonormal basis, then we have
$$
\sum_{i=1}^{n}\mu_{i}\leq\sum_{i=1}^{n}\lambda_{i}\;,\;\; n=1,2,\cdots;\quad\sum_{i}\mu_{i}=\sum_{i}\lambda_{i}.
$$
In our discussion, $P$ is a positive compact operator with eigenvalues $(c_{0}^{2}\geq c_{1}^{2}\geq \cdots)$,
 the Fock basis is an orthonormal basis and $\mathrm{diag}(P)=(\langle 0\vert P\vert 0\rangle\geq\langle 1\vert P\vert 1\rangle\geq\cdots)$. Hence, by the Schur-Horn theorem we have
\begin{equation}\label{e12f}
\langle 0\vert P\vert 0\rangle\leq c_{0}^{2}\;,\; \langle 0\vert P\vert 0\rangle+\langle 1\vert P\vert 1\rangle\leq c_{0}^{2}+c_{1}^{2}\;,\; \cdots \;,\; \sum_{N=0}^{\infty}\langle N\vert
P\vert N\rangle=\sum_{N=0}^{\infty}c_{N}^{2}=1.
\end{equation}
To reach the final result, we rewrite the right hand side of the inequality (\ref{e12d}) as follows:
$$
\begin{array}{c}
  \left(\sum_{N=1}^{\infty}\langle N\vert
P\vert N\rangle-\sum_{N=1}^{\infty}c_{N}^{2}\right)+\left(\sum_{N=2}^{\infty}\langle N\vert
P\vert N\rangle-\sum_{N=2}^{\infty}c_{N}^{2}\right)+ \cdots \\
  =\left(c_{0}^{2}-\langle 0\vert
P\vert 0\rangle\right)+\left(c_{0}^{2}+c_{1}^{2}-(\langle 0\vert
P\vert 0\rangle+\langle 1\vert
P\vert 1\rangle)\right)+ \cdots.
\end{array}
$$
By Eq. (\ref{e12f}), all terms in the brackets are non-negative and this gives rise to the non-negativity of the right hand side of inequality (\ref{e12d}). In this way, we obtain the required result
$$ \Delta(\psi) \geq Z \geq\delta(c).$$
\par
Lemma 1 indicates that for a given set of Schmidt coefficients
$c\,$, that is, for a given amount of entanglement, EPR-like
uncertainties are minimized if the Schmidt vectors are Fock states
in the right order, i.e.
 $ \vert u_{N} \rangle=\vert v_{N} \rangle =\vert N \rangle$ for all $N$.
\par
{\bf{Lemma 2}:} For a given $\Delta\in[b,1)$ and any sequence $
c\,$ with $\Vert c\Vert=1$ and $ \delta(c)=\Delta $, we have $
e(c)\geq e(c^{(\Delta)})$, where $ c^{(\Delta)}$ is the unique
geometric sequence with $\Vert c^{(\Delta)}\Vert=1$ and $
\delta(c^{(\Delta)})=\Delta $.
\par
\textbf{Proof:} As in \cite{8}, the method of Lagrange multipliers is used.
The Lagrange functional is
\begin{equation} \label{L2}
F(c,\lambda,\mu):=e(c)+\frac{\lambda}{2 \ln (2)}[\delta(c)-\Delta]+\frac{\mu +1}{\ln (2)}(\Vert c\Vert -1),
\end{equation}
where $\frac{\lambda}{2 \ln (2)}$ and $\frac{\mu +1}{\ln (2)}$ are positive Lagrange multipliers designed
to simplify the subsequent expressions. Putting from
(\ref{e4a}) and (\ref{e10}) and setting to zero the derivative of Lagrange functional $F(c,\lambda,\mu)$
with repect to $c_{N}$, we obtain
\begin{equation}\label{L3}
c_{N}[N \lambda +\mu -\ln(c_{N}^{2})]=\frac{\lambda
}{a^{2}+\frac{1}{a^{2}}}\;[Nc_{N-1}+(N+1)c_{N+1}],
\end{equation}
where we have defined $c_{-1}=1$. It is clear from Eq. (\ref{L3})
that $c_{N}\neq 0$. Thus we can divide Eq. (\ref{L3}) by $ c_{N} $
and then subtract the same expression with $N$ replaced by $ N+1
$. Defining $
x_{N}:=\frac{c_{N+1}}{\frac{2}{a^{2}+1/a^{2}}\;c_{N}}\;
:=\exp(-2\kappa_{N})\in(0,1] $ for $ N=0,1,... $, and
$\frac{2}{a^{2}+\frac{1}{a^{2}}}:=e^{-2\beta}$, we find
\begin{equation} \label{L4}
x_{N+1}=x_{N}-(A_{N}+B_{N}),
\end{equation}
where
\begin{equation}\label{L5}
A_{N}=\frac{4e^{4\beta}}{N+2}\left[
e^{-2\beta}\sinh^{2}(\kappa_{N}+\beta)-e^{-\beta}\sinh\beta-\frac{2}{\lambda}(\kappa_{N}+\beta)\right],
\end{equation}
\begin{equation}\label{L6}
B_{N}=\frac{Ne^{4\beta}}{N+2}\left[ \frac{1}{x_{N}}-\frac{1}{x_{N-1}}\right].
\end{equation}
Clearly $B_{0}=0$ and the value of $A_{0}$ is fixed for given
values of $\lambda>0$ and $x_{0}$. There exist three
possibilities: (i) $A_{0}>0$. Then, Eq. (\ref{L4}) gives
$x_{1}=x_{0}-A_{0}<x_{0}$ and hence $\kappa_{1}>\kappa_{0}$. This
yields $B_{1}>0$. We now show that $A_{0}>0$ also requires that
$A_{1}>0$. When $A_{0}>0$, then we have
\begin{equation}\label{L6a}
\frac{2}{\lambda}<\frac{e^{-2\beta}\sinh^{2}(\kappa_{0}+\beta)}{\kappa_{0}+\beta}-\frac{e^{-\beta}\sinh\beta }{\kappa_{0}+\beta},
\end{equation}
and therefore $A_{1}$ satisfies
$$
\frac{3A_{1}}{4e^{4\beta}(\kappa_{1}+\beta)}=\left[e^{-2\beta}\frac{\sinh^{2}(\kappa_{1}+\beta)}{\kappa_{1}+\beta}
-\frac{e^{-\beta}\sinh\beta}{\kappa_{1}+\beta}-\frac{2}{\lambda}\right]
$$
$$
>\left[e^{-2\beta}\left(\frac{\sinh^{2}(\kappa_{1}+\beta)}{\kappa_{1}+\beta}-
\frac{\sinh^{2}(\kappa_{0}+\beta)}{\kappa_{0}+\beta}\right)
+e^{-\beta}(\frac{1}{\kappa_{0}+\beta}-\frac{1}{\kappa_{1}+\beta})\sinh\beta\right],
$$
where in the last line we utilized Eq. (\ref{L6a}). It can easily
be checked that for $\kappa_{1}>\kappa_{0}$, the right hand side
of the above inequality is positive and hence $A_{1}>0$.
Therefore, $x_{2}=x_{1}-(A_{1}+B_{1})<x_{1}$. In this manner, we
conclude that $x_{N}$ is decreasing. So $x_{N}$ can achieve a
negative value for some finite N, which is impossible. (ii)
$A_{0}<0$. Then a similar argument as in the first case shows that
$x_{N}$ is increasing. So the normalization condition of c cannot
be fulfilled. Hence, the only possibility is that (iii) $A_{0}=0$.
This leads to the equality of all $x_{N}$ and hence all
$\kappa_{N}$. Denoting the common values of $\kappa_{N}$ by
$\alpha$, we have $x_{N}=e^{-2\alpha}$ and hence by iteration
$c_{N}=c_{0}e^{-2N(\alpha+\beta)}$ for all $N$. To make $A_{0}=0$,
we have to take the Lagrange multiplier $\lambda$ to be
$$\lambda=\frac{2e^{2\beta}(\alpha+\beta)}{\sinh^{2}(\alpha+\beta)-e^{\beta}\sinh\beta}\;.$$
By applying  $
\sum_{N=0}^{\infty} c_{N}^{2}=1 $, we get
\begin{equation} \label{d1}
  c_{0}^{2}=1-e^{-4(\alpha+\beta)}.
\end{equation}
Putting the expression of $ c_{N}$ in Eq. (\ref{e10}), gives
\begin{equation} \label{d2}
  \Delta=\delta(c)=1-\frac{2(1-e^{2\alpha})}{1-e^{4(\alpha+\beta)}}.
\end{equation}
On the other hand, we know that the same value of
$\Delta=\delta(c)$ can also be achieved by the two-mode squeezed
state $|\psi_{r_{\Delta'}}\rangle$ with $\Delta'=\delta'(c)$.
Therefore, by Eq. (\ref{e7a}) we can also write
\begin{equation} \label{d2a}
\Delta=\Delta(\psi_{r_{\Delta'}})=\cosh(2r_{\Delta'})-\frac{2}{a^{2}+\frac{1}{a^{2}}}\sinh(2r_{\Delta'}).
\end{equation}
By equating two expressions (\ref{d2}) and (\ref{d2a}) of
$\Delta$, we obtain
\begin{equation} \label{d2b}
e^{-2(\alpha+\beta)}=\tanh (r_{\Delta'}).
\end{equation}
Hence, Eq. (\ref{d1}) gives $ c_{0}=\frac{1}{\cosh r_{\Delta'}}$.
Finally, we have
$$c_{N}^{(\Delta)}=c_{0}e^{-2N(\alpha+\beta)}=\frac{\tanh^{N}r_{\Delta'}}{\cosh r_{\Delta'}}.$$
As we have mentioned before and it is also clear from Figure 1, there exist two different $r_{\Delta'}$ which
give rise to the same value of $\Delta$. To decide which of them minimizes $e(c)$, let us calculate
$e(c^{(\Delta)})$ as
$$
e(c^{(\Delta)})=-\ln(1-x)-x\ln x\;,\;\;x:=\exp[-4(\alpha+\beta)].
$$
Using the facts that the function $e(c^{(\Delta)})$ is increasing in $x$ and the smaller value of $r_{\Delta'}$
corresponds to the smaller value of $x$ by Eq. (\ref{d2b}), we conclude that for a given $\Delta$, the smaller
 $r_{\Delta'}$ minimizes $e(c)$ and it is the desired solution.
Therefore, among states with the same amount of EPR-like
uncertainty, the squeezed state $|\psi_{r_{\Delta'}}\rangle$ with
smaller $r_{\Delta'}$ has minimal entanglement.
\par
Finally, with the help of Lemmas 1 and 2, Proposition 1 can be
proved as in \cite{8}. However, we include their proof here for
completeness.
\par
\textbf{Proof of proposition 1 :} Given a two-mode state
$|\psi\rangle$, the proposition is trivial in the case of
$\Delta(\psi)=1$. Since, in this case Eq. (\ref{e8}) gives
$r_{\Delta'(\psi)}=0$ and hence by Eq. (\ref{e5}) we have
$E(\psi_{r_{\Delta'(\psi)}=0})=0$. For the case of
$\Delta(\psi)\in[b,1)$, using Eq. (\ref{e4a}) and Lemma 2, we have
$$E(\psi)=e(c)\geq e(c^{(\Delta)})=E(\psi_{r_{\delta'(c)}}).$$
Furthermore, it follows from Lemma 1 and Eq. (\ref{e8}) that
$r_{\Delta'(\psi)}\leq r_{\delta'(c)}$. As $E(\psi_{r})$ increases
in a monotonic manner with $r$, we have
$$E(\psi_{r_{\delta'(c)}})\geq E(\psi_{r_{\Delta'(\psi)}})$$
which completes the proof.
\par
So far, we have shown that for a given value of the EPR-like
uncertainty in the range $[b,1)$, the pure two-mode squeezed state
with smaller squeezing parameter has minimal entanglement among
all two-mode pure states, or equivalently, that for a given amount
of entanglement, the pure two-mode squeezed state has minimal
EPR-like uncertainty. This equivalence follows from Lemma 1 and
the fact that for a pure two-mode squeezed state, the entanglement
and EPR-like uncertainty behave quite oppositely versus squeezing
parameter. Consequently, for a TMGS $\sigma$ with the EPR-like
uncertainty $\Delta(\sigma)$ given by Eq. (\ref{efa}), it is
expected and will be proved in Proposition 2 that the optimal
pure-state decomposition would consist of the standard two-mode
squeezed state with squeezing parameter $r_{\Delta'(\sigma)}$ and
all of its displaced versions
\begin{equation}\label{z1}
\sigma= \int d\xi g(\xi)
\hat{D}(\xi)\vert\psi_{r_{\Delta'(\sigma)}}\rangle\langle\psi_{r_{\Delta'(\sigma)}}\vert
\hat{D}^{\dagger}(\xi).
\end{equation}
Here $g(\xi)$ is a weight function. To calculate $g(\xi)$, we
multiply both sides of Eq. (\ref{z1}) by $\hat{D}(\eta)$ and then
take the trace
\begin{equation}\label{z2}
 {\mathrm{Tr}} \left( \hat{D}(\eta)\sigma\right) = \int d\xi g(\xi) {\mathrm{Tr}} \left( \hat{D}(\eta) \hat{D}(\xi)
 \vert\psi_{r_{\Delta'(\sigma)}}\rangle\langle\psi_{r_{\Delta'(\sigma)}}\vert \hat{D}^{\dagger}(\xi)\right).
\end{equation}
Using the identity $
\hat{D}(\eta)\hat{D}(\xi)=\hat{D}(\xi)\hat{D}(\eta)\exp\left(
-i\eta^{T}\Omega\xi\right)$
 and the cyclic property of trace operation, we have
\begin{equation}\label{k1}
{\mathrm{Tr}} \left( \hat{D}(\eta)\sigma\right)=\int d\xi
g(\xi)\exp\left( -i\eta^{T}\Omega\xi\right)
{\mathrm{Tr}}(\hat{D}(\eta)\vert\psi_{r_{\Delta'(\sigma)}}\rangle\langle\psi_{r_{\Delta'(\sigma)}}\vert).
\end{equation}
Inserting
$${\mathrm{Tr}} \left( \hat{D}(\eta)\sigma\right)=\chi_{\sigma}(\eta)=\exp(-\frac{1}{4}\eta^{T}\Omega^{T}\gamma_{\sigma}\Omega\eta),$$
and
$${\mathrm{Tr}}(\hat{D}(\eta)\vert\psi_{r_{\Delta'(\sigma)}}\rangle\langle\psi_{r_{\Delta'(\sigma)}}\vert)=
\chi_{\psi_{r_{\Delta'(\sigma)}}}(\eta)=\exp(-\frac{1}{4}\eta^{T}\Omega^{T}\gamma_{\psi_{r_{\Delta'(\sigma)}}}\Omega\eta)$$
we get
\begin{equation}\label{k1a}
\exp\left[-\frac{1}{4}\eta^{T}\Omega^{T}(\gamma_{\sigma}-\gamma_{\psi_{r_{\Delta'(\sigma)}}})\Omega\eta\right]=\int
d\xi g(\xi)\exp(-i\eta^{T}\Omega\xi),
\end{equation}
in which
$\gamma_{\sigma}-\gamma_{\psi_{r_{\Delta'(\sigma)}}}\geq0$.
Multiplying both sides by $ \exp(i\eta^{T}\Omega\xi^{\prime})$,
taking the integral over $\eta$ and using the following integral
identities \cite{13}
\begin{equation}\label{k3}
\begin{array}{c}
 \int d^{2n}\lambda \exp  (-\frac{1}{2}\lambda^{T}Q\lambda+i
\lambda^{T}x)=\frac{(2\pi)^{n}\exp
(-\frac{1}{2}x^{T}Q^{-1}x)}{\sqrt{\det(Q)}}, \\
 \hspace{.3cm} \int d^{2n}\eta \exp\left[i\eta^{T}\Omega(\xi-\xi^{\prime})\right]=(2\pi)^{2n}\delta^{2n}(\xi-\xi^{\prime}),\\
\end{array}
\end{equation}
with $ Q $ a real positive definite symmetric matrix, finally $
g(\xi)$ is calculated to be
\begin{equation}\label{k4}
g(\xi)=\frac{1}{\pi^{2}\sqrt{\det(\gamma_{\sigma}-\gamma_{\psi_{r_{\Delta'(\sigma)}}})}}
\exp\left[-\xi^{T}(\gamma_{\sigma}-\gamma_{\psi_{r_{\Delta'(\sigma)}}})^{-1}\xi\right].
\end{equation}
\par
Since $\hat{D}(\xi)$ are local unitary operators, the average
entanglement of the decomposition (\ref{z1}) is equal to
$E[\psi_{r_{\Delta'(\sigma)}}]$. Introducing the auxiliary
function $f:\; (b,1]\rightarrow [0,\infty)$
\begin{equation}\label{e21}
    f(\Delta)=c_{+}(\Delta)\log_{2}[c_{+}(\Delta)]-c_{-}(\Delta)\log_{2}[c_{-}(\Delta)]
\end{equation}
as in \cite{8}, where $c_{\pm}(\Delta)=\frac{(\Delta^{-1/2}\pm\Delta^{1/2})^{2}}{4}$ and
$f$ is a convex and decreasing function of $\Delta$, we can write
\begin{equation}\label{e22}
    E(\psi_{r_{\Delta'}})=f(\Delta').
\end{equation}
\par
Up to now, the parameter $a$ was assumed to be arbitrary. As
mentioned before, for the special value $a=-a_{0}$ the inequality
(\ref{e1}) is a necessary and sufficient condition for the
separability of any TMGS. Hence, for this value we have
$\Delta(\sigma)<1$ iff the TMGS $\sigma$ is entangled. From now
on, let us set $a=-a_{0}$ and denote by $\Delta_{0}(\sigma)$ the
EPR-like uncertainty of an entangled TMGS $\sigma$ for this value.
Then, from Eq. (\ref{efa}) we have
\begin{equation}\label{efaa}
\Delta_{0}(\sigma)=
\frac{a_{0}^{2}\frac{nr_{1}+n/r_{1}}{2}+\frac{mr_{2}+m/r_{2}}{2a_{0}^{2}}-(\sqrt{r_{1}r_{2}}k_{x}
-k_{p}/\sqrt{r_{1}r_{2}})}{a_{0}^{2}+\frac{1}{a_{0}^{2}}}.
\end{equation}
\par
{\bf{Proposition 2:}} Let $\sigma$ be a TMGS with EPR-like
uncertainty $\Delta_{0}(\sigma)\in[b_{0},1)$ where
$\Delta_{0}(\sigma)$ is given by Eq. (\ref{efaa}) and
$b_{0}:=\sqrt{1-\frac{4}{(a_{0}^{2}+\frac{1}{a_{0}^{2}})^{2}}}$.
Then, we have
\begin{equation}\label{e23}
    E_{F}(\sigma)=f(\Delta'_{0}(\sigma)),
\end{equation}
where
\begin{equation}\label{e24}
  \Delta'_{0}(\sigma)=\frac{\Delta_{0}(\sigma)+\sqrt{\Delta_{0}^{2}(\sigma)-b_{0}^{2}}}{1+\sqrt{1-b_{0}^{2}}}.
\end{equation}
The proof of Proposition 2 is the same as the one presented in
\cite{8}. However, we restate the proof for completeness.
\par
\textbf{Proof:} Let $D$ be an arbitrary pure-state decomposition
of $\sigma$ as $\sigma=\sum_{k} p_{k}
|\psi_{k}\rangle\langle\psi_{k}|$ and $D_{0}$ be its decomposition
given by Eq. (\ref{z1}). Average entanglements of the
decompositions $D$ and $D_{0}$ are $\bar{E}(D)=\sum_{k} p_{k}
E(\psi_{k})$ and
$\bar{E}(D_{0})=E(\psi_{r_{\Delta'_{0}(\sigma)}})=f(\Delta'_{0}(\sigma))$,
respectively. Now it is enough to prove that $\bar{E}(D)\geq
f(\Delta'_{0}(\sigma))$ for any decomposition $D$ since the
decomposition $D_{0}$ already saturates this bound. As a
consequence of Lemma 1 and Eq. (\ref{e22}), we have
$$E(\psi_{k})\geq E(\psi_{r_{\Delta'_{0}(\psi_{k})}})=f(\Delta'_{0}(\psi_{k})).$$
So, we can write
$$\bar{E}(D)\geq\sum_{k} p_{k}f(\Delta'_{0}(\psi_{k}))\geq f(\sum_{k}p_{k}\Delta'_{0}(\psi_{k})),$$
where the second inequality follows from convexity of the function
$f$. As a simple result of the definition of EPR-like uncertainty by Eq. (\ref{e2}),
we have the inequality $\Delta_{0}(\sigma)\geq
\sum_{k}p_{k}\Delta_{0}(\psi_{k})$ and hence $\Delta'_{0}(\sigma)\geq
\sum_{k}p_{k}\Delta'_{0}(\psi_{k})$. The latter inequality together with
the fact that $f$ is a decreasing function of its argument give
$f(\sum_{k}p_{k}\Delta'_{0}(\psi_{k}))\geq f(\Delta'_{0}(\sigma))$ which completes the proof.
\section{Discussions and examples}
\textbf{a.} \textbf{Symmetric TMGS}
\par
For an entangled symmetric TMGS $\tilde{\sigma}$, we have $n=m$
and $k_{x}\geq -k_{p}>0$. Setting these values in Eqs. (\ref{ca})
and (\ref{cb}) and solving them for $r_{1}$ and $r_{2}$,  give
$$r_{1}=r_{2}=\sqrt{\frac{n+k_{p}}{n-k_{x}}}.$$
In this case, we obtain $a_{0}=1$ and hence $b_{0}=0$. Using these
values, Eqs. (\ref{efaa}) and (\ref{e24}) yield
$$\Delta'_{0}(\tilde{\sigma})=\Delta_{0}(\tilde{\sigma})=\sqrt{(n-k_{x})(n+k_{p})}.$$
Finally, Eq. (\ref{e23}) gives
$$E_{F}(\tilde{\sigma})=f\left(\sqrt{(n-k_{x})(n+k_{p})}\right)$$
which is in agreement with the result of the Giedke et al.'s work
\cite{8}.
\par
As a confirmation of the result (\ref{e23}), it is now an
opportunity to make another proof for the Theorem 1 of \cite{14}
based on this result.
\par
\textbf{Theorem 1 of \cite{14}:} The EOF of any TMGS with a given
EPR-like uncertainty is greater than or equal to the EOF of a
mixed symmetric TMGS with the same EPR-like uncertainty.
\par
\textbf{Proof:} Let $\sigma$ be an arbitrary TMGS with EPR-like
uncertainty $\Delta_{0}(\sigma)$ and $\tilde{\sigma}$ be a
symmetric TMGS with the same EPR-like uncertainty
$\Delta_{0}(\tilde{\sigma})=\Delta_{0}(\sigma)$. Using the fact
that the function $f$ is a decreasing function of its argument,
the problem is reduced to the verification of the validity of
inequality $\Delta'_{0}(\tilde{\sigma})\geq\Delta'_{0}(\sigma)$.
We have
$\Delta'_{0}(\tilde{\sigma})=\Delta_{0}(\tilde{\sigma})=\Delta_{0}(\sigma)$.
Setting this in Eq. (\ref{e23}), gives
$$\Delta'_{0}(\sigma)=\frac{\Delta'_{0}(\tilde{\sigma})
+\sqrt{{\Delta'}_{0}^{2}(\tilde{\sigma})-b_{0}^{2}}}{1+\sqrt{1-b_{0}^{2}}}.$$
Rewriting this equation as
$$\Delta'_{0}(\sigma)=\Delta'_{0}(\tilde{\sigma})
\frac{1+\sqrt{1-\frac{b_{0}^{2}}{{\Delta'}_{0}^{2}(\tilde{\sigma})}}}{1+\sqrt{1-b_{0}^{2}}},$$
and using the fact that ${\Delta'}_{0}^{2}(\tilde{\sigma})\leq 1$,
verify
the validity of the inequality and hence complete the proof.\\\\
\textbf{b.} \textbf{Two-mode squeezed thermal states}
\par
The class of two-mode squeezed thermal states $\check{\sigma}$ is
a significant class of TMGSs. For such a state, we have $n\geq m$ and
$k_{x}=-k_{p}>0$. In this case, solving Eqs. (\ref{ca}) and
(\ref{cb}) gives $r_{1}=r_{2}=1$. With these values, we obtain
$b_{0}=\frac{n-m}{n+m-2}$ and Eqs. (\ref{efaa}) and (\ref{e24})
yield
$$\Delta_{0}(\check{\sigma})=\frac{n\tilde{m}+m\tilde{n}-2k_{x}\sqrt{\tilde{n}\tilde{m}}}{\tilde{n}+\tilde{m}},$$
$$\Delta'_{0}(\check{\sigma})=\left[\frac{\sqrt{n\tilde{m}-k_{x}\sqrt{\tilde{n}\tilde{m}}}
+\sqrt{m\tilde{n}-k_{x}\sqrt{\tilde{n}\tilde{m}}}}{\sqrt{\tilde{n}}+\sqrt{\tilde{m}}}\right]^{2},$$
where we have made the definitions $\tilde{n}:=n-1$ and
$\tilde{m}:=m-1$.
\par
In \cite{16}, Giovannetti et al. determined the EOF for a family
of two-mode thermal states with parameters
$n=2(\bar{n}+1)\kappa-1$, $m=2(\bar{n}+1)\kappa-(2\bar{n}+1)$ and
$k_{x}=-k_{p}=2(\bar{n}+1)\sqrt{\kappa(\kappa-1)}$ where
$\kappa\in[1,\infty)$ is the gain parameter of an amplifier
channel and $\bar{n}\in[0,\infty)$ is the average photon number.
Their EOF is independent of $\bar{n}$ and its value is given by
$g(\kappa):=\kappa\log_{2}[\kappa]-(\kappa-1)\log_{2}[\kappa-1]$.
However, our method shows that for a given value of $\kappa$, the
EOF is $\bar{n}$ dependent, its values are always less than $g(\kappa)$ and approaches $g(\kappa)$ as $\bar{n}$ gets large.\\\\
\textbf{c.} \small{\textbf{Our method versus the Gaussian EOF and
the lower and upper bounds of EOF}}
\par
The Gaussian EOF was introduced by Wolf et al. \cite{17} as a version
of the EOF for bipartite Gaussian states in which only
decompositions into pure Gaussian states are considered. In the
case of a symmetric TMGS, this measure coincides with the exact
EOF while for a general TMGS it provides an upper bound. For a
general TMGS $\rho$ with standard-form parameters $(n, m, k_{x},
k_{p})$, if we define
$$C_{x}:=
\left(
\begin{array}{cc}
n & k_{x}\\
k_{x} & n\\
\end{array}
\right)\quad,\quad\;C_{p}:= \left(
\begin{array}{cc}
m & k_{p}\\
k_{p} & m\\
\end{array}
\right),$$ then, as shown in \cite{17, 18}, its Gaussian EOF
$E_{GF}(\rho)$ is given by minimum value of the entropy of
entanglement of a pure gaussian state $\rho_{_{P}}$ whose CM is of
the form $\gamma_{_{P}}=\Gamma\oplus\Gamma^{-1}$. Here, $\Gamma$
is a real positive $2\times 2$ matrix as
$$\Gamma=\left(
\begin{array}{cc}
x_{0}+x_{3} & x_{1}\\
x_{1} & x_{0}-x_{3}\\
\end{array}
\right),$$ where $x_{0}, x_{1}$ and $x_{3}$ are real parameters
satisfying
\begin{equation}\label{det}
    \mathrm{det}(C_{x}-\Gamma)=\mathrm{det}(\Gamma-C_{p}^{-1})=0.
\end{equation}
The reduced covariance matrix $\gamma_{_{P}}^{(A)}$ of the state $\rho_{_{P}}$ is
\begin{equation}\label{det1}
\gamma_{_{P}}^{(A)}=\left(
\begin{array}{cc}
x_{0}+x_{3} & 0\\
0 & (x_{0}-x_{3})/\mathrm{det}\Gamma\\
\end{array}
\right).
\end{equation}
It was found that the Gaussian EOF $E_{GF}(\rho)$ is given by
\begin{equation}\label{gef}
    E_{GF}(\rho)=f[({m_{opt}})^{1/2}-(m_{opt}-1)^{1/2}],
\end{equation}
in which $f$ is the function defined by Eq. (\ref{e21}) and
$m_{opt}$ is the minimum value of the determinant of $\gamma_{_{P}}^{(A)}$:
$$\mathrm{det}\gamma_{_{P}}^{(A)}=1+\frac{x_{1}^{2}}{\mathrm{det}\Gamma}.$$
To obtain $m_{opt}$, it is enough to minimize
$\mathrm{det}\gamma_{_{P}}^{(A)}$ under the constraints of Eq.
(\ref{det}).
\par
In the work \cite{14}, Rigolin et al have derived a lower bound on
the EOF of a general mixed TMGS with CM as in Eq. (\ref{v2}). This
lower bound is the EOF of a symmetric TMGS $\tilde{\sigma}$ whose
standard-form parameters are
$$n_{\tilde{\sigma}}=m_{\tilde{\sigma}}=\frac{n+m}{2},\quad (k_{x})_{\tilde{\sigma}}=k_{x},\quad(k_{p})_{\tilde{\sigma}}=k_{p},$$ $$(r_{1})_{\tilde{\sigma}}=(r_{2})_{\tilde{\sigma}}=\left[\frac{(n+m)/2+k_{p}}{(n+m)/2-k_{x}}\right]^{1/2}.$$
Oliveira et al. \cite{15} established a tight upper bound for the
EOF of a general TMGS employing the necessary properties of
Gaussian channels. This upper bound is the EOF of a symmetric TMGS
$\breve{\sigma}$ provided that a matrix with parameters
$$n_{\breve{\sigma}}=m_{\breve{\sigma}}=m,\quad (k_{x})_{\breve{\sigma}}=k_{x},\quad(k_{p})_{\breve{\sigma}}=k_{p},$$ $$(r_{1})_{\breve{\sigma}}=(r_{2})_{\breve{\sigma}}=\sqrt{\frac{n+k_{p}}{n-k_{x}}}.$$
constitutes a bona fide CM for $\breve{\sigma}$.
Also they argued that the EOF of a general TMGS $\rho$ with CM as in Eq. (\ref{v2}), must satisfy
\begin{equation}\label{}
E_{F}(\tilde{\sigma})\leq E_{F}(\rho)\leq E_{F}(\breve{\sigma})
\end{equation}
Also, Marians \cite{10} developed an approach for evaluating the exact
EOF of a general TMGS.
\par
In the Table 1, we give upper and lower
bounds together with Gaussian EOF and the exact EOFs computed
based on Marians and our approach for the six mixed TMGSs given in
\cite{14}.
\begin{table}[h]
\renewcommand{\arraystretch}{1.2}
\addtolength{\arraycolsep}{6pt}
$$
\scriptsize{\begin{array}{|c|c|c|c|c|c|}\hline \mathrm{n, m, k_x,
k_p} & E_{F}(\tilde{\sigma}) & \mathrm{Marians \ E_{F}(\rho)} &
\mathrm{our \ E_{F}(\rho)} & E_{F}(\breve{\sigma})&
\mathrm{E_{GF}(\rho)} \\ \hline\hline
  \mathrm{2, 1.5, 1.2, -1} & \mathrm{0.28919} & \mathrm{0.3836537397} & \mathrm{0.3784745926} & \mathrm{--}& \mathrm{0.3836537389}\\
   \hline
   \mathrm{2, 1.5, 1, -1} & \mathrm{0.14672} & \mathrm{0.2027415462} & \mathrm{0.2022298409} & \mathrm{0.56616}& \mathrm{0.2027415477}\\
   \hline
   \mathrm{3, 2, 1.8, -1.2} & \mathrm{0.00681} & \mathrm{0.04851229950} & \mathrm{0.04850819279} & \mathrm{--}& \mathrm{0.04851230013}\\
   \hline
   \mathrm{2.6, 1.7, 1.3, -0.9} & \mathrm{0} & \mathrm{0.01198094416} & \mathrm{0.01198079698} & \mathrm{0.40946}& \mathrm{0.01198094462}\\
   \hline
   \mathrm{3,2, 1.7, -1.2} & \mathrm{0.00142} & \mathrm{0.01398144359} & \mathrm{0.01398132663} & \mathrm{--}& \mathrm{0.01398144137}\\
   \hline
   \mathrm{2.5, 2, 1.3, -1.2} & \mathrm{0.00001} & \mathrm{0.002510512206} & \mathrm{0.002510511701} & \mathrm{0.14838}& \mathrm{0.002510512809}\\
   \hline
  \end{array}}
$$
\caption{\footnotesize{The first column shows the parameters of
the CM in its standard form. The next columns show the values of
$E_{F}(\tilde{\sigma})$ given by \cite{14}, $E_{F}(\rho)$ based on Marians and our
approach, $E_{F}(\breve{\sigma})$ and Gaussian EOF $E_{GF}(\rho)$,
respectively. States on rows 1, 3 and 5 do not give a
physical state $\breve{\sigma}$.}}
\renewcommand{\arraystretch}{1}
\addtolength{\arraycolsep}{3pt}
\end{table}
As implied by the table, our values of $E_{F}(\rho)$ fall inside the valid region
between upper and lower bounds and
are obviously less than Marians values. Also, our values are less than the values of
$E_{GF}(\rho)$ as they should be.

\section{Conclusion}
We have defined an EPR-like uncertainty by using Duan et al.'s
inequality and showed that among pure states with a given amount
of entanglement, pure squeezed states have the least entanglement.
Then, for a TMGS we attained the optimal pure-state decomposition
which leads to its EOF and provided a simple method for the
evaluation of the EOF. For the two special and important cases of
symmetric TMGSs and squeezed thermal states we have determined the
EOF explicitly. We expect that this work will provide new insight
into the subject of Gaussian states entanglement.
\par {\bf Acknowledgment:} We acknowledge very valuable discussions with M. A. Jafarizadeh
and also thanks M. A. Fasihi for interesting and useful
discussions. This research was supported by a research fund No.
401.22 from Azarbaijan Shahid Madani University.


\begin{thebibliography}{10}
\bibitem{1}
Charles H. Bennett, Gilles Brassard, Sandu Popescu, Benjamin
Schumacher, John A. Smolin, and William K. Wootters, {\it Phys.
Rev. Lett.} {\bf 76}, 722 (1996).
\bibitem{2}
William K. Wootters, {\it Phys. Rev. Lett.} {\bf 80} 2245 (1998).
\bibitem{3}
V. Vedral and M. B. Plenio, {\it Phys. Rev. A} {\bf 57}, 1619
(1998).
\bibitem{3a}
M. B. Plenio and S. Virmani, {\it Quantum Inf. Comput.} {\bf 7}, 1
(2007).
\bibitem{4}
Charles H. Bennett, David P. DiVincenzo, John A. Smolin, and
William K. Wootters, {\it Phys. Rev. A} {\bf 54}, 3824 (1996).
\bibitem{5}
A. Ferraro, S. Olivares, and M. G. A. Paris, {\it Gaussian States
in Quantum Information} (Bibliopolis, Napoli, 2005).
\bibitem{6}
J. Eisert and M. B. Plenio, {\it Int. J. Quantum Inform.} {\bf 1},
479 (2003).
\bibitem{7}
S. L. Braunstein and P. Van Loock, {\it Rev. Mod. Phys.} {\bf 77},
513 (2005).
\bibitem{8}
G. Giedke, M. M. Wolf, O. Kr\"{u}ger, R. F. Werner, and J. I.
Cirac, {\it  Phys. Rev. Lett.} {\bf 91}, 107901 (2003).
\bibitem{17}
M. M. Wolf, G. Giedke, O. Kr\"{u}ger, R. F. Werner, and J. I.
Cirac, {\it Phys. Rev. A} {\bf 69}, 052320 (2004).
\bibitem{18}
G. Adesso and F. Illuminati, {\it Phys. Rev. A} {\bf 72}, 032334
(2005).
\bibitem{9}
J. Solomon Ivan and R. Simon, arXiv: 0808.1658v1.
\bibitem{10}
Paulina Marian and Tudor A. Marian,  {\it Phys. Rev. Lett.} {\bf
101},  220403 (2008).
\bibitem{14}
G. Rigolin and C.O. Escobar, {\it Phys. Rev. A} {\bf 69}, 012307
(2004).
\bibitem{15}
F. Nicacio and M. C. de Oliveira, {\it Phys. Rev. A} {\bf 89}, 012336 (2014).
\bibitem{16}
V. Giovannetti, R. Garcia-Patron, N. J. Cerf, and A.S. Holevo,
{\it Nature Photonics} {\bf 8}, 796-800 (2014); arXiv:1312.6225.
\bibitem{11}
Lu-Ming Duan, G. Giedke, J. I. Cirac, and P. Zoller, {\it Phys.
Rev. Lett.} {\bf 84}, 2722 (2000).
\bibitem{12}
R. Simon, {\it Phys. Rev. Lett.} {\bf 84}, 2726 (2000).
\bibitem{19}
William Arveson, Richard V. Kadison, {\it Contemporary
Mathematics}, {\bf 414} no.1, 247-263 (2006).
\bibitem{20}
Victor Kaftal, Gary Weiss, {\it J. Funct. Anal.}, {\bf 259} no.
12, 3115-3162 (2010).
\bibitem{13}
S. Olivares, {\it Eur. Phys. J. Special Topics}, {\bf 203}, 3
(2012).
\end{thebibliography}
\end{document}